\documentclass[preprint,review,authoryear,12pt]{elsarticle}

\usepackage{amssymb}
\usepackage{graphicx, natbib, stmaryrd, supertabular}
\journal{Expert Systems with Applications}

\begin{document}

\begin{frontmatter}
\let\today\relax



\title{Towards a Satisfactory Conversion of Messages among Agent-based Information Systems}


\author{Idoia Berges\corref{cor1}}\ead{idoia.berges@ehu.es}
\author{Jes{\'u}s Berm{\'u}dez}\ead{jesus.bermudez@ehu.es}
\author{Alfredo Go{\~n}i}\ead{alfredo@ehu.es}
\author{Arantza Illarramendi}\ead{a.illarramendi@ehu.es}
\cortext[cor1]{Corresponding author. Phone:+34 943 01 81 51}

\address{University of the Basque Country, UPV/EHU. Paseo Manuel de Lardizabal, 1, Donostia-San Sebasti{\'a}n, Spain}

\begin{abstract}
Over the last years, there has been a change of perspective concerning the management of information systems, since they are no longer isolated and need to communicate with others. However, from a semantic point of view, real communication is difficult to achieve due to the heterogeneity of the systems. We present a proposal which, considering information systems are represented by software agents, provides a framework that favours a semantic communication among them, overcoming the heterogeneity of their agent communication languages. The main components of the framework are a suite of ontologies --conceptualizing communication acts-- that will be used for generating the communication conversion, and an Event Calculus interpretation of the communications, which will be used for formalizing the notion of a satisfactory conversion. Moreover, we present a motivating example in order to complete the explanation of the whole picture.
\end{abstract}


\end{frontmatter}

% \linenumbers
\newtheorem{definition}{Definition}
\section{Introduction}\label{Introduction}
Nowadays, the vision of isolated information systems that work on their own without cooperating with other systems is no longer realistic. Information systems which have been independently developed by different organizations need to communicate with each other in order to enhance their functionality. However, several problems must be solved before real communication is achieved. Although there are several levels where communication issues may occur, such as the transport level or the application protocol level, in this paper we focus only on how to achieve utterance communication at a semantic level. 
%This latter problem is widely recognized and goes far beyond the use of XML for the interchange of data --because even if it has been proved relevant for this issue, it does not deal with the semantics of the exchanged data.

We adopt the cooperative software agents approach as communication platform since agent technology has been proved useful for solving problems with a highly distributed nature that need flexible and adaptable solutions (e.g.\ \citep{Agogino12, Sato11}).
Communication between software agents is based on the interchange of messages. When an agent from one system wants to communicate with an agent of another system, it generates a message following the rules that have been established in the former system in order to write messages. These rules imply the use of fixed structures, languages and meaning, which is useful within a certain agent system but becomes useless in the majority of situations when trying to communicate between agents from heterogeneous systems. We advocate the use of a more flexible interoperation where the interpretation of a message is made dynamically and transparently for the heterogeneous agents that want to communicate with each other. 

In the relevant literature, several works can be found that consider the problem of how to achieve a flexible communication between agents from heterogeneous systems. Details of this are presented in Section \ref{Related} but, in summary, we can say that a high number of solutions point towards the development of mechanisms that only deal with syntactical aspects of agent communication languages (ACLs) (see \citep{Lopes05}, \citep{Suguri02}). 
Furthermore, a trend of works emphasize the adequacy of communication protocols in order to restrict the potentially wild set of utterance choices. Models of human dialogues have been used to categorize some types of dialogues \citep{Walton95} and protocol designers have adopted such source as a basis for structuring agent communications \citep{Maudet02}. Analyses have been undertaken on desiderata for such protocols \citep{McBurney02}. 
Doubtless, communication protocols are an important part of the play of software agent interactions, and they are subject to a lot of study and effort \citep{Desai09, Berges11, Chopra09}. Nevertheless, heterogeneity of already existing information systems exhibits great differences in the structure and intended semantics of their messages. Therefore it is not guaranteed that they satisfy the basic assumptions considered in some protocol models (i.e. semantics of individual messages expected in the protocol do not match with the message semantics of the participating agent). 
In our opinion, semantic technologies such as ontologies can be used in order to handle semantic aspects properly, and in particular, we advocate for managing semantics of individual messages in order to increase the possibilities of success when dealing with protocols. Other authors share the opinion that the introduction of semantic-based technology in information systems provides for a significant enhancement of their functionalities and capabilities. For example authors in \citep{Shaw12} study the application of ontologies in order to create FAQ auto-categorizing systems that can improve customer service in technical support centers and alike. Moreover, in \citep{Chen12} ontologies and the Semantic Web Rule Language (SWRL\footnote{www.w3.org/Submission/SWRL/}) are used in order to develop a diabetes medication recommendation system. This system aims at suggesting a prescription for a patient based on knowledge about symptoms and diabetes-related drugs. Finally authors in \citep{Belmonte08} describe an ontology-based multiagent decision support system for bus fleet management, whose goal is to reason about traffic behaviour in the same way an expert traffic operator would. However, as opposed to our approach, only agents that use the same ACL are considered. In this sense, we present a framework consisting of a communication acts ontology, called \textsc{CommOnt}, a translation mediator, an Event Calculus axiomatization of the scenario and some agents in charge of managing the communication process between systems.
% Other works also promote the use of Semantic Web technologies in the considered scenario (see \citep{Pasha06}, \citep{Willmott05b}), however, as far as we know they are in a less mature state than our proposal. 

In summary, the main features of the proposed framework are:
\begin{itemize}
\item It favours a semantic communication between heterogeneous and distributed information systems using semantic technologies and therefore it increases their cooperation opportunities.
\item It incorporates an ontology that describes the types of communication acts used by software agents. This ontology allows the recognition of instances of communication acts from one system as instances of communication acts of another system.
\item It also integrates  domain and action ontologies with the goal of extending its applicability to a wide range of different scenarios.
\item It incorporates Event Calculus sentences to represent the scenario. Model theory is used to formalize a notion of satisfactory conversion. Event Calculus reasoners allow to check proposed conversions.
\item A developed prototype allows us to observe the behaviour of the framework in specific scenarios. In particular, in this paper, the feasibility of our framework is presented in a scenario where one agent of our \textsc{MedicalFIPAAgents} system communicates with one agent of the \textsc{Aingeru} system, two healthcare information systems which use different languages, structures and meanings in the composition of their messages.
\end{itemize}

The rest of the paper is organized as follows: In Section 2 the main components of the proposed framework are presented, as well as a brief description of the communication process. Section 3 explains preliminary notions about the ontologies in the framework and the dynamic interpretation of communication acts. 
Section 4 introduces the formalization of the notion of satisfactory conversion, while one scenario of the proposed framework at work is presented in Section 5. Related works are discussed in Section 6 and finally, we end with some conclusions in Section 7.

\section{Global architecture of the framework}\label{Architecture}
In this section the architecture of the framework is presented from a global point of view. First, the elements that belong to the framework are briefly described, emphasizing the role that each element plays in the architecture. Then, the process applied to one message from the moment it is created until it reaches its final destination is shown.

\subsection{Elements of the framework}\label{Architecture:Elements}

\begin{figure}
\begin{center}	
	\includegraphics[width=4.4in]{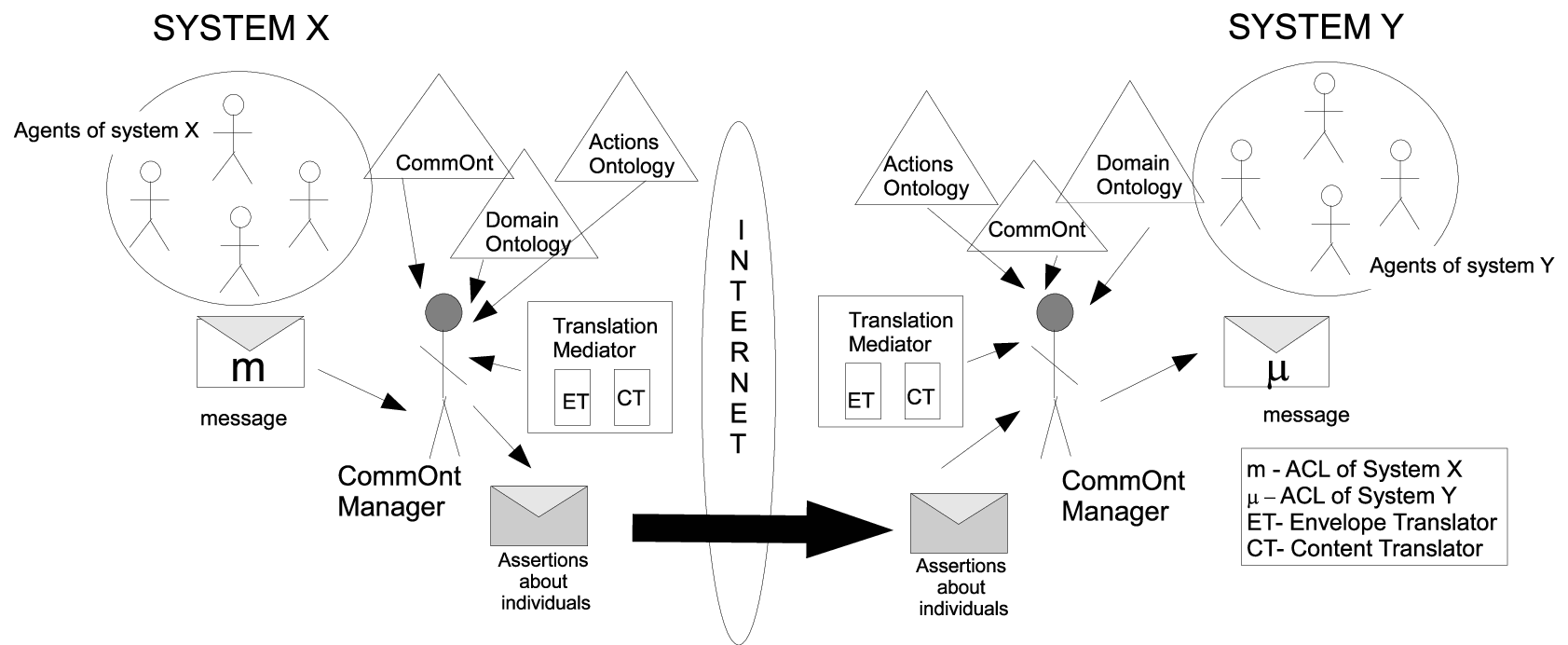} %architecture
	\caption{\label{architectureFig} Architecture of the framework}
\end{center}
\end{figure}

In Fig.~\ref{architectureFig} the global architecture of the framework can be found. The following elements take part in it:

\vspace{0.2cm}
\textbf{\textsc{CommOnt} and other ontologies}: We have defined the \textsc{CommOnt} ontology to serve as a conceptualization of communication acts.  Further details about this ontology will be explained in Section \ref{Ontologies}. Moreover, a set of ontologies formalize the conceptualization of domain actions and domain propositions.

\vspace{0.2cm}
\textbf{TranslationMediator}: This module is in charge of the translators that transform the messages managed by the agents of each particular information system to their representation in terms of the \textsc{CommOnt} ontology and the other way round. In other words, translators considered in our framework produce an OWL \citep{OWL2} document from an input message written 
in a given ACL and conversely, produce an ACL message given an OWL document as input. 

\vspace{0.2cm}
\textbf{CommOntManager:} This agent is responsible for managing the process applied to a message from its creation by the sender until it reaches the receiver. In order to do this the CommOntManager needs to interact with the TranslationMediator and the ontologies. The state machine which represents the behaviour of the CommOntManagers is shown in Appendix \ref{appendix:stateMachines}. We are aware that the role of the CommOntManager could be supressed from the framework, but this would imply that all the agents that belong to the information systems were able to interact with the TranslationMediators and the ontologies, so we think that it is more convenient to have a single agent specialized in that task.

\subsection{Brief description of the conversion process}\label{Architecture:Process}
%In Fig X the procces applied to a message sent by an agent of system X to an agent of system Y can be found.
\begin{figure}
\begin{center}	
	\includegraphics[width=4.4in]{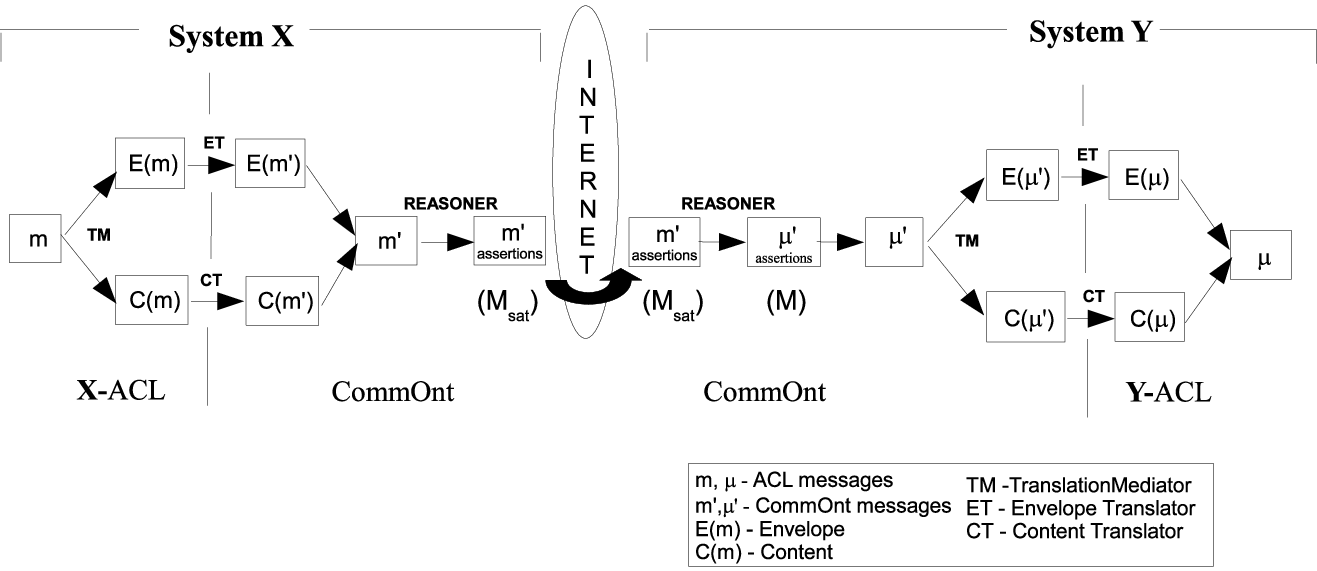}
	\caption{\label{messageProcess} Conversion process applied to a message}
\end{center}
\end{figure}

When an agent of information system $X$ (see Fig.~\ref{architectureFig}) wants to communicate with an agent of system $Y$, first it must send the message to the CommOntManager of its system. Then the following process is carried out (see Fig.~\ref{messageProcess}): In the first step, the CommOntManager sends the message \textsf{m} to the TranslationMediator module, which translates it to the terms (i.e. classes and properties) in the \textsc{CommOnt} ontology. In order to do so, it first divides the message into two submessages, the envelope \textsf{E(m)} and the content \textsf{C(m)} and sends each of the submessages to its corresponding translator, \textsf{ET} and \textsf{CT} respectively. The envelope of a message contains the intention of the message; for example, to make a request or to give information about something, and also some information about the communication process, such as the identity of the sender and the receiver. The content of the message contains the object of the intention; for instance, what is the specific request made or which information is given.
%An envelope translator will be used to translate the envelope of the message from the agent communication language of the information system $X$ to the terms in \textsc{CommOnt} and simultaneously, a content translator will be used to represent the content of the message from the content language of the information system $X$ with the content language classes in \textsc{CommOnt}. 
In general, for every ACL and for every content language there will exist a translator between those languages and \textsc{CommOnt}, and vice versa.

Once both submessages have been transformed to \textsc{CommOnt} sentences, they are re-joined. The resultant message can be considered a set of basic OWL2 assertional axioms (i.e. Description logic ABox assertions of the type \textit{C(x)}, \textit{R(x,y)}, where \textit{C} is a class name, \textit{R} is a property name, \textit{x} is a named individual, and \textit{y} is either a named individual or a literal value). Then, an OWL2 reasoner is launched to calculate the set of basic assertions (technically known as ABox realization) derived from the previous set of assertions in the context of ontologies concerning the sender system $X$.

More formally, let $M$ be the set of basic ABox assertions (i.e. of the form  \textit{C(x)}, \textit{R(x,y)} with \textit{C}, \textit{R}, \textit{x} and \textit{y} described as before) from the original message, and $O^X$ the set of ontologies concerning the system $X$ (including its version of \textsc{CommOnt}, and domain and action ontologies). Then we call $M_{der}$ the set of new basic ABox assertions that can be derived from $M$ in the context of $O^X$ (($O^X, M)$$\vdash$$M_{der}$).

At this point, the saturated set $M_{sat}$=$M$$\cup$$M_{der}$ is materialized as OWL2 assertional axioms and sent to the CommOntManager of the information system $Y$. Again, an OWL2 reasoner is launched in the context of the corresponding $O^Y$ to derive basic ABox assertions from $M_{sat}$. As consequence, new information is inferred, which relates the terms of the original message, that were only understood by agents of system $X$, with the terms in the \textsc{CommOnt} ontology of system $Y$, which are understood by the agents of system $Y$. Then, the new information is transformed into an OWL-format message and sent to the TranslationMediator module, which divides it into two submessages (envelope and content) and sends each one to its corresponding translator, in order to translate both parts from \textsc{CommOnt} to the languages used in system $Y$. After the translations, the resulting submessages are joined to form a complete message understood by agents of system $Y$. The CommOntManager can then send the message to its intended receiver, and the conversion process is finished.

\subsubsection*{Discussion}
Choices made for the previously explained conversion process deserve some comments. First of all, a common formalism is selected for representing the 
diversity of ACLs. We chose the OWL2 language, which is based on a logic formalism 
(i.e., an expressive Description logic), and which was carefully designed for describing concepts, mantaining a trade-off between rich expressivity and decidable computational properties. OWL2 has exhibited enough expressivity in 
various application domains and we found it appropriate for specifying faceted classes of messages. OWL2 ontologies are logical systems that define a set of 
axioms that enable automated reasoning over a set of given facts. OWL2 ontology 
development allows a much more decentralized and collaborative approach than 
other non logic-based proposals.
Ontology axioms corresponding to the \textit{application layer} capture the 
structure and some semantics of the actual agent communication language. 
Application layer development is an ontology design task that must be carried 
out by domain experts from each information system's administration staff. 
Descriptions of classes and properties of messages can be checked for semantic 
relationships within the logic model. Translation from the actual agent 
communication language to the ontology language and vice versa can be made by any 
appropriate language translation process. We are not primarily concerned with 
that translation process.

Our approach presents a communication acts ontology which describes implementation independent classes and properties of messages, and which 
can include descriptions of established standards for ACLs. That ontology 
will be a \textit{common layer} shared by all participating information systems.   
Ontologists who design the application layer of each information system 
must develop alignment axioms among classes and properties from the 
application layer and the common layer that will serve as an articulation ontology.  Those alignment axions are the glue of the conversion process.

An individual message from one information system will be eventually converted 
in a message for another information system by an inferencing process guided by 
the alignment axioms connecting both application layers to the articulation 
common layer. Some advantages of the inferencing approach against a syntax 
transformation approach may be considered. For instance, in case of $n$ classes 
$A_1 \ldots A_n$ from the application layer such that 
$\forall i=1..n\ A_i\sqsubseteq C$, for some class $C$ in the common layer, 
$n$ syntax translation processes should be defined in order to cope with 
any message from $A_1 \ldots A_n$, in contrast with the simple entailment step which derives $m \in C$ from $m \in A_i$ and $A_i\sqsubseteq C$. Also, a 
class $A$ in the application layer may be declared as a subclass of two different 
classes $C$ and $D$ in the common layer (i.e., $A\sqsubseteq C \sqcap D$); 
then, using the syntax transformation approach, an \textit{ad hoc} class should be added to the common layer in order to capture such a kind of messages, 
with the undesirable effect of garbage generation, in contrast with the 
inference based approach, which has no trouble coping with the entailments 
$m \in C$ and $m \in D$. Moreover, the syntax transformation approach should 
cope with a confluence problem in cases such as $A\sqsubseteq C$, 
$C\sqsubseteq D$, with $A$ in the application layer and $C$, $D$ in the common layer. Since $A$ is subclass of $C$ and $D$, two transformation process $T_C$ 
and $T_D$ must be defined; another translation process $P$ should be defined due to $C\sqsubseteq D$, could it be guaranteed that composition of $T_C$ with $P$ is 
equivalent to $T_D$? In summary, the inference process approach exhibits better 
performance due to the underlying logic properties.

Another interesting aspect of the conversion process we want to stress is the 
need for the materialization of the entailed basic assertions. It is important 
for the process to make explicit those assertions that describe properties 
of the original message with terms from the common layer; since, once sent 
to the target information system, they will be in charge of conveying 
the semantics (notice that assertions with particular application layer terms 
will not be understood in the recipient information system).

\section{Preliminaries} \label{Preliminaries}
In this section preliminary notions about the ontologies in the framework and our adopted approach for the dynamic interpretation of communication acts are explained.
\subsection{Ontologies}\label{Ontologies}
\begin{figure}
\begin{center}	
	\includegraphics[width=4.8in]{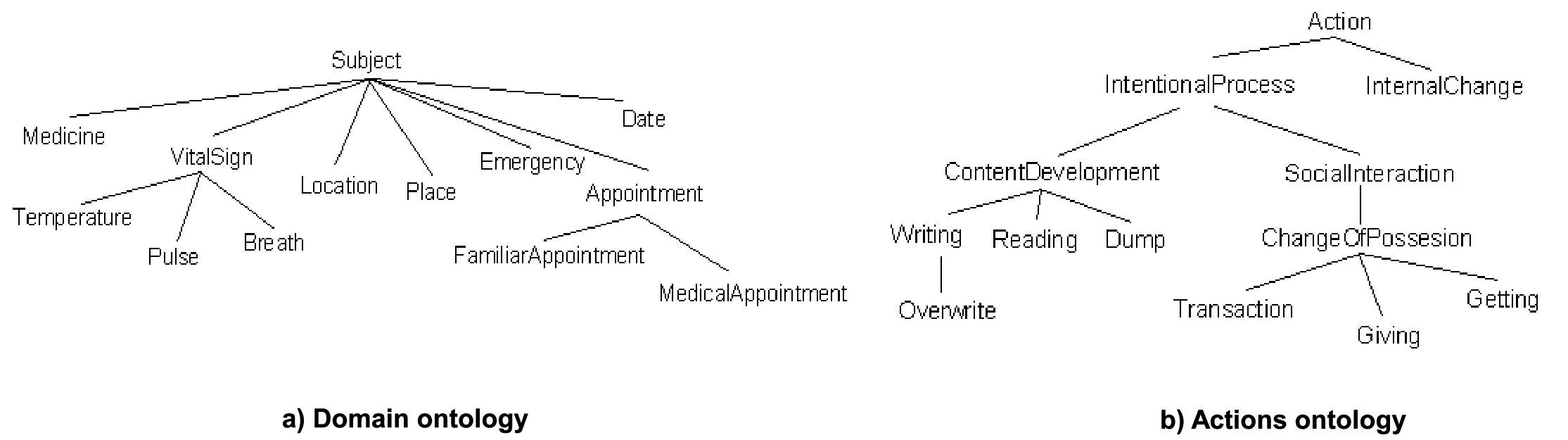}
	\caption{\label{domainAndActionsOntologies} Domain and action ontologies}
\end{center}
\end{figure}

As stated in the previous subsection, the developed framework makes use of different kinds of ontologies. Domain ontologies are needed for describing classes that refer to entities, properties or facts from specific knowledge areas. The terms of the domain ontologies are used, among others, in the content of messages. In Fig.~\ref{domainAndActionsOntologies}a a subset of the domain ontology used in the \textsc{Aingeru} system can be found. Action ontologies are useful for describing actions that agents can perform. Classes such as \texttt{Reading}, \texttt{Writing}, \texttt{Dump}, etc. can be found in such ontologies. As it happens with domain ontologies, the terms from the action ontologies are used in the content of messages. In Fig.~\ref{domainAndActionsOntologies}b a subset of the well known SUMO \citep{SUMO} ontology is shown. Although domain and action ontologies are also important, the distinguishing ontology of our framework is a formal specification of a conceptualization of communication acts, which we have named \textsc{CommOnt}.

The main goal of the \textsc{CommOnt} ontology is to facilitate semantic communication between heterogeneous agent-based information systems. The core of the ontology is formed by the so called communication acts. The main design criteria adopted for describing communication acts has been to follow the \textit{speech
acts} theory originated by Austin (\citep{Austin62}), further developed by Searle (\citep{Searle69}) and elaborated by Vanderveken (\citep{Vanderveken90}). The speech acts theory is a linguistic theory that has been recognized as the principal source of inspiration for designing most well known standard agent communication languages. In the following section we explain briefly the specification of \textsc{CommOnt}, preceded by the conceptualization principles that drive that specification. 

\subsubsection{\textsc{CommOnt} ontology: Conceptualization of communication acts}

According to Austin's theory every communication act may be viewed as the sender's expression of an
attitude toward some possibly complex proposition. A sender performs a
communication act which is expressed by a coded message and is
directed to a receiver. Therefore, a communication act has two main
components.  First, the attitude of the sender which is called the
\textit{illocutionary force} (\textit{F}), that expresses social
interactions such as informing, requesting or promising, among
others (those appear in the envelope part of the message, as we mentioned in Section \ref{Architecture:Process}). And second, the \textit{propositional content} (\textit{p})
which is the subject of what the attitude is about.
That perspective has been called the \textit{F(p) framework}.

According to Vanderveken, the notion of illocutionary force is not taken as a primitive notion, but can be composed of six dimensions: an illocutionary point, a mode of achievement, propositional content conditions, preparatory conditions, sincerity conditions and a degree of strength.
There are five basic illocutionary points, and for each of them there is a counterpart primitive illocutionary force which has that point: \textit{Assertive}, \textit{Commissive}, \textit{Directive}, \textit{Declarative}, and \textit{Expressive}, respectively.
The common features of those primitive forces are that they have one illocutionary point, no special mode of achievement of that point and a neutral degree of strength.
The remaining illocutionary forces are derived from the five primitive by applying a finite number of operations: the addition of a certain propositional content condition $\alpha_1$ (indicated as $\phi_{\alpha_1}$), the addition of a certain preparatory condition $\alpha_2$ ($\sigma_{\alpha_2}$), the addition of a certain sincerity condition $\alpha_3$ ($\psi_{\alpha_3}$), the restriction of the mode of achievement $\alpha_4$ ($\mu_{\alpha_4}$) and increasing ($+_n$, $n\geq 1$) or decreasing ($-_n$, $n\geq 1$) the degree of strength.

For example, departing from the primitive directive force \textit{Directive}, the following forces can be obtained:
\begin{itemize}
\item \textit{Request} = $\mu_{\alpha}$(\textit{Directive}), where $\alpha$=``polite".
\item \textit{Urge} = ($+_1\ \circ\ \sigma_{\alpha}$)(\textit{Request}), where $\alpha$=``The speaker has reasons for that course of actions"
%\item \textit{TellTo} = $\mu_{\alpha}$(\textit{Directive}), where $\alpha$=``peremptory".
%\item \textit{Demand} = $+_1$(\textit{TellTo}).
%\item \textit{Require} = $\sigma_{\alpha}$(\textit{Demand}), where $\alpha$=``The propositional content is compulsory".
\end{itemize}

The aforementioned operations can be composed functionally, and moreover, their composition is commutative ($f\circ g$ = $g\circ f$).

\subsubsection{\textsc{CommOnt} ontology: Specification}
In \textsc{CommOnt}  the \textit{F(p)} framework is used and 
different kinds of illocutionary forces and contents
leading to different classes of communication acts are supported.
\textsc{CommOnt} contains one main category to represent the communication acts. There are also two other categories to represent the actors that take part in the communication and the content of the messages, in which classes from domain and action ontologies are integrated. 
The terms of the \textsc{CommOnt} ontology are described in the form of classes and properties using the Web Ontology Language OWL2. \textsc{CommOnt} is composed of two interrelated layers (common layer and applications layer) that classify the communication acts regarding different levels of abstraction, being the common layer the most general and the applications layer the most specific. This division into layers allows a clearer visualization of the ontology, but it does not imply a technical division of the ontology. Every information system will have its own version of the \textsc{CommOnt} ontology. The common layer will be the same in the \textsc{CommOnt} ontology of all the systems. However, the applications layer will be tailored to each system. 

The \textit{common layer} of the \textsc{CommOnt} ontology includes the top class \texttt{Com\-mu\-ni\-ca\-tion\-Act}, which represents the universal class of messages\footnote{For the presentation we prefer a logic notation instead of the more verbose \textsc{rdf/\textsc{xml}} syntax.}. Every concrete instance of a message is an individual of this class.
\begin{figure}
\begin{center}	
	\includegraphics[width=4.8in]{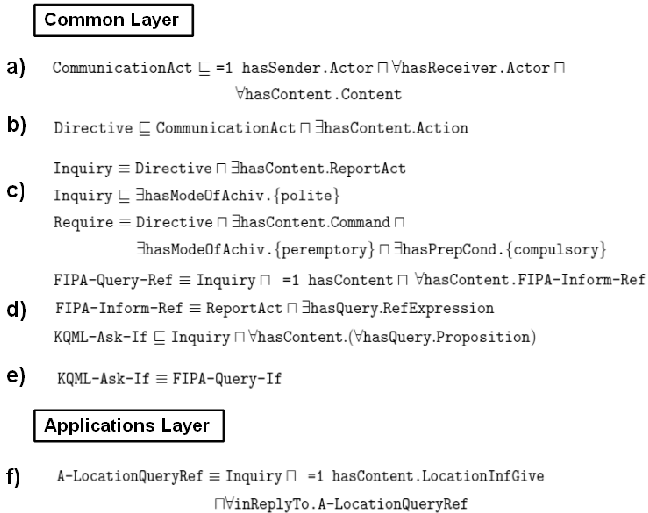}
	\caption{\label{commont} Some axioms of the \textsc{CommOnt} ontology}
\end{center}
\end{figure}
In addition, five general subclasses of \texttt{CommunicationAct} are defined, each of which refers to a primitive illocutionary point: \textit{Assertive}, \textit{Directive}, \textit{Commissive}, \textit{Expressive} 
and \textit{Declarative} (Fig.~\ref{commont}b). Finally, subclasses of those five primitive communication acts have been defined (Fig.~\ref{commont}c). Due to Vanderveken's conceptualization, membership to the primitive classes is not inferred, but asserted (that is the reason why subclass axioms are used for their specification). However, their subclasses can be specified by necessary and sufficient conditions, so that their membership can be inferred based on properties of the message.

Moreover, the common layer is extended with specific classes of messages that belong to different Agent Communication Languages (notice that different ACL standards use distinct classes, Fig.~\ref{commont}d). More precisely, communication acts of the standards FIPA (\citep{FIPA-ACL}) and KQML (\citep{KQML}) are considered. Also, alignment axioms between communication acts from different standard languages are defined in this level, so that communication between software agents that use different standard languages is feasible (Fig.~\ref{commont}e).

The \textit{applications layer} is the system specific part, where communication acts that are particular of each information system are described. These communication acts constitute the particular agent communication language of a system. Moreover, they are defined as subclasses of the communication acts in the common layer of the \textsc{CommOnt} ontology. For example, one of the classes that belongs to the applications layer of the \textsc{CommOnt} ontology of the \textsc{Aingeru} system is the one in Fig.~\ref{commont}f.

In order for an information system to benefit from the interoperability that is provided by the proposed framework, its administrator must make the effort of defining the communication acts used by that system in the applications layer of its version of the \textsc{CommOnt} ontology. However, notice that in case that the system is adhered to one of the described standards, there is no need to redefine those communication acts in the applications layer. Moreover, it is obvious that the more alignment axioms are defined with the terms of the common layer, the easier it will be to achieve communication between different information systems.

\subsection{Dynamic interpretation of communication acts}
In addition to the static representation of the communication acts in an ontology, we have defined a dynamic interpretation of them. More precisely, we have adopted a first-order logic interpretation of messages, and then, model-theory serves as the foundation of that notion. Since messages are taken as actions (i.e. communication acts), they can be distinguised by their effects, which allows us to talk about a dynamic interpretation of communication acts. The following subsections present briefly the logical foundations of our approach.

\subsubsection{Event Calculus}\label{sec:eventCalculus}

% 1: Argumentamos que vamos a usar un FOL como referencia canónica para la interpretación de los mensajes. Decimos que vamos a usar el Event Calculus

% 2: Presentamos el EC (en particular la axiom. DEC). Nociones sigma, etc

The Event Calculus \citep{Shanahan99} offers a language (based on first-order logic) to reason about actions and their effects. Its basic ontology comprises \textit{events}, \textit{fluents} and \textit{timepoints}: \textit{events} 
correspond to \textit{actions} in our context; \textit{fluents} are predicates whose truth 
value may change over time. 
The Event Calculus contains several predicates about events, fluents and timepoints which are summarised in Table \ref{table:ECpredicates}. See \citep{Mueller06} for the complete set of predicates.

\begin{table}
\centering
\begin{tabular}{ll}
\hline
\multicolumn{1}{c}{\textbf{Predicate}} & \multicolumn{1}{c}{\textbf{Textual description}}\\
\hline
\textbf{\textit{Happens(e,t)}} & Event \textit{e} happens or occurs at timepoint \textit{t}.\\
\textbf{\textit{HoldsAt(f,t)}} & Fluent \textit{f} is true at timepoint \textit{t}.\\
\textbf{\textit{ReleasedAt(f,t)}} & Fluent \textit{f} is released from the commonsense law of\\ & inertia at timepoint \textit{t}.\\
\textbf{\textit{Initiates(e,f,t)}} & Event \textit{e} initiates fluent \textit{f} at timepoint \textit{t}.\\
\textbf{\textit{Terminates(e,f,t)}} & Event \textit{e} terminates fluent \textit{f} at timepoint \textit{t}.\\
\textbf{\textit{Releases(e,f,t)}} & Event \textit{e} releases fluent \textit{f} at timepoint \textit{t}.\\
\hline
\end{tabular}
\caption{\label{table:ECpredicates}Some predicates of the Event Calculus}
\end{table}

The Event Calculus uses linear time. In \citep{Mueller06} a Discrete Event Calculus axiomatization (\texttt{DEC}) of its featuring predicates can be found, which restricts the timepoint sort to discrete time steps (which is adequate for our context). Table \ref{table:DECaxioms} shows some of the axioms in \texttt{DEC} (see the aforementioned reference for the complete set).

\begin{table}
\centering
\begin{tabular}{lr}
\hline
\multicolumn{1}{c}{\textbf{Axiom and textual description}} & \multicolumn{1}{c}{\textbf{Tag}}\\
\hline
\textit{HoldsAt(f,t+1)} $\leftarrow$ \textit{HoldsAt(f,t)} $\wedge$ $\neg$\textit{ReleasedAt(f,t+1)} $\wedge$ & (\texttt{\textbf{DEC5}})\\
\hspace{1in}$\neg\exists$\textit{e} (\textit{Happens(e,t)} $\wedge$ \textit{Terminates(e,f,t)})\\
If a fluent is true at timepoint \textit{t}, the fluent is not released from the\\
commonsense law of inertia at timepoint \textit{t+1}, and the fluent is not\\ 
terminated by any event that occurs at \textit{t}, then the fluent is true at \textit{t+1}.\\  
\hline
 \textit{HoldsAt(f,t+1)} $\leftarrow$  \textit{Happens(e,t)} $\wedge$ \textit{Initiates(e,f,t)}   &(\texttt{\textbf{DEC9}})\\
If a fluent is initiated by some event that occurs at timepoint \textit{t},\\
then the fluent is true at \textit{t+1}.\\
\hline
 $\neg$\textit{HoldsAt(f,t+1)} $\leftarrow$  \textit{Happens(e,t)} $\wedge$ \textit{Terminates(e,f,t)}  &(\texttt{\textbf{DEC10}})\\
If a fluent is terminated by some event that occurs at timepoint \textit{t},\\
then the fluent is false at \textit{t+1}.\\
%\hline
% \textit{ReleasedAt(f,t+1)} $\leftarrow$  \textit{Happens(e,t)} $\wedge$ \textit{Releases(e,f,t)}  &(\texttt{\textbf{DEC11}})\\
%If a fluent is released by some event that occurs at timepoint \textit{t},\\
%then the fluent is released from the commonsense law of intertia at \textit{t+1}.\\
%\hline
% $\neg$\textit{ReleasedAt(f,t+1)} $\leftarrow$  \textit{Happens(e,t)} $\wedge$ (\textit{Initiates(e,f,t)} $\vee$ \textit{Terminates(e,f,t)})  &(\texttt{\textbf{DEC12}})\\
%If a fluent is initiated or terminated by some event that occurs at\\ 
%timepoint \textit{t}, then the fluent is not released from the commonsense\\  
%law of intertia at \textit{t+1}.\\  
\hline
\end{tabular}
\caption{\label{table:DECaxioms}Subset of axioms of the Discrete Event Calculus that depict the influence of events on fluents.}
\end{table}

The Event Calculus was initially developed for commonsense reasoning. A suitable set of dedicated formulas must be asserted in order to axiomatize the domain of interest. Such formulas can be grouped into intuitive collections (See \citep{Mueller06} for additional explanations):

\begin{itemize}
\item \textit{Observations} [$\Gamma$] are formulas of the form \textit{HoldsAt(f,t)} and \textit{ReleasedAt(f,t)}.
\item A \textit{narrative} [$\Delta$] is a set of event ocurrence formulas and temporal ordering formulas. \textit{Happens(e,t)} is an \textit{event ocurrence formula}.
\item \textit{Effect axioms} [$\Sigma$] are a set of formulas of the form $\gamma$$\Rightarrow$\textit{Initiates(e,f,t)} or
$\gamma$$\Rightarrow$\textit{Terminates(e,f,t)}, where $\gamma$ is a context condition.
\item \textit{State constraints and Event ocurrence constraints} [$\Psi$] are a set of formulas of the form $\gamma_1$, $\gamma_1\Rightarrow\gamma_2$ or $\gamma_1\Leftrightarrow\gamma_2$, and \textit{Happens($e_1$,t)}$\wedge\gamma_1\Rightarrow$($\neg$)\textit{Happens($e_2$,t)}, respectively, where $\gamma_1$ and $\gamma_2$ are context conditions.

\end{itemize}

%\begin{definition}
%Let $\gamma_1$ and $\gamma_2$ be context conditions, $e_1$ and $e_2$ events, and $t$ a timepoint. Then, $\gamma_1$, $\gamma_1\Rightarrow\gamma_2$ and $\gamma_1\Leftrightarrow\gamma_2$ are \textit{state constraints}, and \textit{Happens($e_1$,t)}$\wedge\gamma_1\Rightarrow$($\neg$)\textit{Happens($e_2$,t)} is an \textit{event ocurrence constraint}.
%.
%\end{definition}
Finally, it is worth mentioning that by default, no unexpected effects or events occur. In the Event Calculus this is ensured by using a technique called Circumscription \citep{McCarthy80}, which minimizes the extension of the predicates of the Event Calculus. The way circumscription (CIRC) is computed is out of the scope of this paper, but again the technical details can be found in \citep{Mueller06}.

There are Event Calculus reasoners that are able to compute the set of observations that are true in timepoint \textit{t+1}, given a set of observations $\Gamma_{t}$ in timepoint \textit{t}, a narrative $\Delta$, effect axioms $\Sigma$ and state and event occurrence constraints $\Psi$. We express this with the following notation:

\texttt{DEC} $\wedge$ CIRC($\Sigma$; \textit{Initiates}, \textit{Terminates}, \textit{Releases}) $\wedge$ CIRC($\Delta$; \textit{Happens}) $\wedge$ $\Gamma_{t}$ $\wedge$ $\Delta$ $\wedge$ $\Sigma$ $\wedge$ $\Psi$ $\vdash_{_{+1}}\Phi_{t+1}(\Sigma,\Delta,\Psi,\Gamma_{t})$

\subsubsection{Commitments}\label{sec:commitments}

A special kind of fluents, relevant to our context, are the so-called commitments, whose notion has been referenced often in the Artificial Intelligence literature \citep{Singh96,Castelfranchi95}. Different kinds of commitments have been distinguished: the psychological (or internal) commitments, the dialogical commitments, and the social commitments. Psychological commitments are established autonomously by an agent, which is committed to his beliefs, desires and intentions only towards himself. Formal semantics based on such mental concepts has been associated to communication acts. However, that option has been 
criticized for its approach \citep{Singh98} as well as for its analytical difficulties 
\citep{Wooldridge00}.
On the other hand, dialogical and social commitments present an external character. Dialogical commitments stem from the work in \citep{Hamblin70} and have had a special influence in the study of agent argumentation dialogues \citep{Walton95}; these are commitments held by participants in a dialogue to support statements they have made if questioned or challenged to do so by other participants. Social commitments \citep{Singh00,Venkatraman99,Fornara03} are established by participation in certain social situations. According to 
the social approach, when agents interact they become involved in social obligations to 
each other, which help in structuring and harmonizing multiagent systems and in attaining coherence in their actions. Those commitments are public, and therefore they are 
suitable for an objective and verifiable semantics of agent interaction. As a result, we have adopted this latter social approach  
to express the dynamic semantics of communication acts described in the \textsc{CommOnt} ontology. 

\begin{definition}
A \textit{base-level commitment} \textsf{C}(\textit{x, y, p}) is a ternary relation that represents 
a commitment made by \textit{x} (the \textit{debtor}) to \textit{y} (the \textit{creditor}) to bring 
about a certain proposition \textit{p}.
\end{definition} 

For example, the base-level commitment \textsf{C}(\textit{Alice, Bob, bidForPainting}) indicates the commitment made by agent Alice to agent Bob to bid for a painting.

Moreover, sometimes an agent accepts a commitment only if a certain condition holds or when a certain commitment is made by another agent. This is called a conditional commitment.

\begin{definition}
A \textit{conditional commitment} \textsf{CC}(\textit{x, y, q, p}) is a quaternary relation 
that represents that if the condition \textit{q} is brought out, \textit{x} will be committed to 
\textit{y} to bring about the proposition \textit{p}.
\end{definition}

For example, \textsf{CC}(\textit{Alice, Bob, pricePainting(\textless 35M\$), bidForPainting}) indicates the commitment made by agent Alice to agent Bob to bid for a painting if the price of the painting is less than 35M\$.

%In our context, commitments (base-level and conditional) can be considered fluents, and semantics of 
%communication acts can be expressed with predicates in the Event Calculus. 

%\begin{definition} 
%A \textit{contextualized fluent} \textsf{F}$^\omega$ is a fluent with an associated context $\omega=<\alpha_{1},\alpha_{2},\alpha_{3},\alpha_{4},n>$, where $\alpha_{1}$ is a mode of achievement, $\alpha_{2}$ is a preparatory condition, $\alpha_{3}$ is a sincerity condition, $\alpha_{4}$ is a propositional content condition, and $n\ (-\infty<n<\infty)$ is a degree of strength.
%\end{definition}

Furthermore, some formulas are needed to capture the dynamics of commitments.
%Commitments 
%are a type of fluent, typically put in force by communication acts, that become inoperative after the 
%appearance of other fluents. 
In the following formulas (taken from \citep{Yolum04}), \textit{e(x)} represents an event caused by 
\textit{x}. The first formula declares that when a debtor of a commitment that is in force causes an event that 
initiates the committed proposition, the commitment ceases to hold.

\textsc{Formula 1:}  
\textit{HoldsAt(}\textsf{C}\textit{(x, y, p), t)} $\wedge$ \textit{Happens(e(x), t)} $\wedge$ \textit{Initiates(e(x), p, t)} $\rightarrow$ 
\textit{Terminates(e(x),}\textsf{C}\textit{(x, y, p), t)}.

The second formula declares that a conditional commitment that is in force disappears and generates a base-level 
commitment when the announced condition is brought out by the creditor.

\textsc{Formula 2:}  
\textit{HoldsAt(}\textsf{CC}\textit{(x, y, c, p), t)} $\wedge$ \textit{Happens(e(y), t)} $\wedge$ \textit{Initiates(e(y), c, t)} $\rightarrow$
\textit{Initiates(e(y),}\textsf{C}\textit{(x, y, p), t)} $\wedge$ 
\textit{Terminates(e(y),}\textsf{CC}\textit{(x, y, c, p), t)}.

Finally, the third formula declares that a conditional commitment disappears when the debtor brings out the committed 
proposition (regardless of the announced condition). 

\textsc{Formula 3:} \textit{HoldsAt(}\textsf{CC}\textit{(x, y, c, p), t)} $\wedge$ \textit{Happens(e(x), t)} $\wedge$ 
\textit{Initiates(e(x), p, t)} $\rightarrow$ \textit{Terminates(e(x),}\textsf{CC}\textit{(x, y, c, p), t)}

\section{Formalization of satisfactory conversion} 

%Once the translation process has finished, it is reasonable to ask whether the communicating agents agree on the understanding of the sent and received message, respectively. It is conceivable that, although their respective message classes are related properly in the communication acts ontology, their intended effects may not match accurately (i.e. the semantics of the sent message is not exactly the same as the semantics of the received message).
%
%In order to formalize a notion of satisfactory translation, we adopt a first-order logic interpretation of messages, and then, model-theory serves as the foundation of that notion. Since messages are taken as actions (i.e. communication acts), they can be distinguised by their effects, which allows us to talk about a dynamic interpretation of communication acts.

%Subsections \ref{sec:eventCalculus} and \ref{sec:commitments} present briefly the logical foundations of our approach. Then, subsection \ref{sec:formalization} presents our formalization of the notion of satisfactory translation, which is the core of the paper.

Once the conversion process has finished, it is reasonable to ask whether the communicating agents agree on the understanding of the sent and received message. It is conceivable that, although their respective message classes are related properly in the communication acts ontology, their intended effects may not match accurately (i.e. the semantics of the sent message is not exactly the same as the semantics of the received message).

In this section, first, the description of the conversion context domain is summarized by presenting an excerpt of a proper axiomatization, and then our definition of a satisfactory conversion is introduced.

\subsection{Description of the conversion context domain}\label{sec:formalization}
The conversion context domain is composed of a narrative, a set of observations, a set of effect axioms and a set of state and event ocurrence constraints.

\subsubsection*{Narrative} 

Let \textit{Happens(send(A(m)), t)} be a sentence for encoding the event of sending one message $m$, which is an instance of the communication act $A$, at timepoint $t$. Since we consider only the conversion of a single message (thus, one event), our narrative is typically a singleton $\Delta_{A(m)}=\{\textit{Happens(send(A(m)), t)}\}$.

\subsubsection*{Observations} 

Let $\Gamma_{t}$ be a set of observations to indicate the fluents that hold at timepoint $t$. Some of these observations are due to the effect axioms (let us call them $\Gamma_\Sigma$), while others are observations to describe the features of the fluents in $\Gamma_\Sigma$. For example, if due to some effect axiom the observation \textit{HoldsAt(P,t)} is obtained, where \textit{P} is of class \textit{ApplicP}, then the observation \textit{HoldsAt(ApplicP(P),t)} indicates that at timepoint \textit{t} it can be said that \textit{P} is of class \textit{ApplicP}. For a specific agent system, the classes and properties that appear in these observations belong to the applications layer of its corresponding \textsc{CommOnt} ontology or to the domain or action ontologies. The information including classes of the common layer (e.g.\ the classes of the common layer obtained by subsumption) is not considered.

\subsubsection*{Effect axioms} 

Formulas 1 to 3 in Section \ref{sec:commitments} are special cases of effect axioms that we group in the set $\Sigma_C$.

Another kind of effect axioms are those which describe the effects associated to the classes of communication acts. We represent them with the set $\Sigma_T$. Next we show some examples that refer to the effects of the communication acts \texttt{Inquiry} and \texttt{Responsive}, 
which appear in the common layer of \textsc{CommOnt}. %poner referencia a la subseccion?
\begin{itemize}
\item 
\textit{Initiates(send(Inquiry(s, r, P))}, \textsf{CC}\textit{(r, s, accept(r, s, P), P), t)}\\ 
By using an Inquiry, the sender expects to get an answer from the receiver to the question indicated in the content of the message. Moreover, the event of sending an Inquiry from \textit{s} to \textit{r} produces the effect of generating a conditional 
commitment which expresses that if the receiver \textit{r} accepts the demand, it will be commited 
to the proposition \textit{P} in the content of the communication act. 
%\item
%\textit{Initiates(send(Require(s, r, P))}, \textsf{CC}\textit{(r, s, accept(r, s, P), P), t)}\\ 
%The event of sending a Require from \textit{s} to \textit{r} produces the effect of generating a conditional 
%commitment which expresses that if the receiver \textit{r} accepts the demand, it will be commited 
%to the proposition \textit{P} in the content of the communication act. 
%\item 
%\textit{Initiates(Accept(s, r, P), accept(s, r, P), t)}.\\
%The sending of an Accept produces the effect of generating the accept fluent.
\item
\textit{Terminates(send(Responsive(s, r, P, RA)),}\textsf{C}\textit{(s, r, RA), t)}.\\
\textit{Terminates(send(Responsive(s, r, P, RA)),}\textsf{CC}\textit{(s, r, accept(s, r, RA), RA), t)}.\\
\textit{Initiates(send(Responsive(s, r, P, RA)),} \textit{P, t)}\\
By using a Responsive, the sender answers a previously received question or demand it committed to answer. As consequence of sending a message of the class Responsive, the commitment (either base-level or conditional) of the sender \textit{s} towards 
the receiver \textit{r} to bring about proposition \textit{RA} ceases to hold, and moreover, the fluent \textit{P} is initiated.
\end{itemize}

\subsubsection*{State and event occurrence constraints} 

Moreover, from the axioms in \textsc{CommOnt}, state and event ocurrence constraints are obtained. We group them in the set $\Psi$.

\newcounter{qcounter}
\begin{list}{\alph{qcounter})}{\usecounter{qcounter}}
\item State constraints: Let $B$ be a class name for representing a communication act. Let $Cexp$ be an expression of the form\footnote{This is just an example of how state constraints are calculated. For other types of $Cexp$ the corresponding rules are created. The whole specification can be found in \citep{Berges11b}.} $D\sqcap\exists p.E$, where $D$, $E$ are class names and $p$ is an object property. Then, two rules for indicating state constraints are the following:
\begin{itemize}
\item if \mbox{$B\sqsubseteq Cexp$} then \textit{HoldsAt(B(?m),t)}$\rightarrow$ \textit{HoldsAt(D(?m),t)}
\item if \mbox{$B\sqsupseteq Cexp$} then \textit{HoldsAt(D(?m),t)} $\wedge$ \textit{HoldsAt(p(?m,?op))} $\wedge$ \textit{HoldsAt(E(?op))} $\rightarrow$ \textit{HoldsAt(B(?m))}
\end{itemize}
\item Event ocurrence constraints: Let $B$ and $C$ be class names for representing communication acts. Then, the following rule can be defined:
\begin{itemize}
\item if \mbox{$B\sqsubseteq C$} then \textit{Happens(send(B(?m)),t)$\rightarrow$ Happens(send(C(?m)),t)}
\end{itemize}
	
\end{list}

In conclusion, the dynamic semantics of a message is determined by the fluents that are initiated and terminated as a result of  
sending that message. 
In summary, communication acts have a dual semantic representation: The description in 
\textsc{CommOnt} of their structure and of their hierarchical relationships and then some Event Calculus formulas which specify their effects. 

\subsection{Definition of satisfactory conversion}

In the following we present the formalization of the notion of a satisfactory conversion of a message. Let $\Gamma_t$ be the set of observations that are true at time $t$, and let $\Sigma$=$\Sigma_C$$\cup$$\Sigma_T$ be a set of effect axioms. Let us assume that message $A(m)$ is sent at time $t$, then we call $\Phi_{t+1}(\Sigma,\Delta_{A(m)},\Psi,\Gamma_{t})$ the set of observations that are true at time $t+1$ in the context of $\Sigma$, $\Psi$ and $\Gamma_t$. Formally:

DEC $\wedge$ CIRC($\Sigma$; \textit{Initiates}, \textit{Terminates}, \textit{Releases}) $\wedge$ CIRC($\Delta_{A(m)}$; \textit{Happens}) $\wedge$ $\Gamma_{t}$ $\wedge$ $\Delta_{A(m)}$ $\wedge$ $\Sigma$ $\wedge$ $\Psi$ $\vdash_{_{+1}}\Phi_{t+1}(\Sigma,\Delta_{A(m)},\Psi,\Gamma_{t})$ 

\begin{definition}

Message A$_{2}(m)$ from system 2 in the context of $\Sigma_2$, $\Psi_2$ and $\Gamma_t$ is a \textit{satisfactory conversion} of message A$_{1}(m)$ from system 1 in the context of $\Sigma_1$, $\Psi_1$ and $\Gamma_{t}$ if $\Phi_{t+1}(\Sigma_{2},\Delta_{A_2(m)},\Psi_{2},\Gamma_{t})$ is consistent and \\
 $\Phi_{t+1}(\Sigma_{2},\Delta_{A_2(m)},\Psi_{2},\Gamma_{t})
\models\Phi_{t+1}(\Sigma_{1},\Delta_{A_1(m)},\Psi_{1},\Gamma_{t})$.
\end{definition}

Notice that $\Phi_{t+1}(\Sigma_{i},\Delta_{A_i(m)},\Psi_{i},\Gamma_{t})$, with both $i=1,\ 2$, is computable and that Event Calculus reasoners are prepared to conduct the desired proof. Intuitively, we consider a message $m'$ a satisfactory conversion of a message $m$ if the effects of $m'$ in the context of the target system are sufficient as effects of $m$ in the context of the source system. That is to say, the intended meaning expressed by the sender (with message $m$) is captured by the receiver (with message $m'$). 
%\stackrel{\Leftrightarrow}{def}

\subsubsection*{Discussion}
One may argue that the logical consequence in Definition 3 should be satisfied in the opposite direction as well (that is, that the effects of $m$ should also entail those of $m'$) in order to consider that a conversion has been satisfactory. However, as stated before, our goal is to check whether the recipient system is able to understand, at least, the intended meaning of the information sent by the sender system. We consider that the recipient system may add some effects that are relevant for the pecularities of that system in such a context. Although this may strengthen meaningfully the effects of the message in that context, it still respects the effects expected by the sender system. As a result we think that placing a two-way logical consequence
would be a too strong condition for communication in open environments
where each agent system chooses its ACL regardless of what others do.

Moreover, some discussion may be raised due to the fact that we take into account a particular initial context $\Gamma_{t}$ at the time of evaluating the correctness of a conversion. It is widely known that the messages exchanged between software agents are not isolated messages, but they are part of conversations regulated by communication protocols. Thus, the same message can occur in the context of multiple conversations. However, we believe that it is not necessary to evaluate the correctness of a conversion for all possible conversations (that is, for every possible initial context $\Gamma_{t}$). If the conversion preserves enough semantics, within the context in use, then the conversion can be considered correct and can be used instead of the original message. It would be unwise to discard a satisfactory conversion done under a context $\Gamma_{t}$ just because it is not correct for some other context $\Gamma'_{t}$.

For example, let us imagine two messages $m_{1}$ and $m_{2}$ which only differ in the class of their communication act: $m_{1}\in A_{1}$ and $m_{2}\in A_{2}$. Let $m_{2}$ be the proposed conversion for $m_{1}$ and suppose that $f$, $g$ and $k$ are fluents. Moreover, for the clarity of the example, let us assume that $\Psi_{1}$ and $\Psi_{2}$ do not play any major role in this case, so they are not described. 

In the following scenario:

\begin{center}
\begin{supertabular}{ll}
\\
\multicolumn{2}{l}{$\Gamma_{t0}$=\{\textit{HoldsAt(f,t0)}\}}\\
\multicolumn{2}{l}{
$\Sigma$=$\Sigma_{C}\cup\Sigma_{1}\cup\Sigma_{2}\cup\Sigma_{3}$ where:}\\
\multicolumn{2}{l}{\hspace{0.3in}$\Sigma_{C}$ is defined as before}\\
\multicolumn{2}{l}{\hspace{0.3in}$\Sigma_{1}$= \textit{Initiates(send($A_1(?m)$),k,t)}}\\
\multicolumn{2}{l}{\hspace{0.3in}$\Sigma_{2}$=\textit{HoldsAt(f,t)}$\Rightarrow$\textit{Initiates(send($A_1(?m)$),g,t)}}\\
\multicolumn{2}{l}{\hspace{0.3in}$\Sigma_{3}$=\textit{Initiates(send($A_2(?m)$),k,t)}}\\
$\Delta_{A_1(m_1)}$=\{\textit{Happens(send($A_1(m_1)$),t0)}\}\\ $\Delta_{A_2(m_2)}$=\{\textit{Happens(send($A_2(m_2)$),t0)}\}\\
\\
\end{supertabular}
\end{center}

In order to prove that $m_2$ can not be considered a satisfactory conversion for $m_1$, first $\Phi_{t0+1}(\Sigma, \Delta_{A_1(m_1)},\Psi_{1},\Gamma_{t0})$ is calculated:

\begin{enumerate}
\item From $\Sigma_{1}$,  $m_{1}\in A_{1}$, and due to the binding of the variables taking into account the current scenario: \textit{Initiates(send($A_1(m_1)$),k,t0)} 
\item From $\Gamma_{t0}$,  $m_{1}\in A_{1}$, and $\Sigma_{2}$, and due to the binding of the variables considering the current scenario: \textit{Initiates(send($A_1(m_1)$),g,t0)}
\item From $\Delta_{A_1(m_1)}$, 1, 2 and \texttt{DEC9} (see Table \ref{table:DECaxioms}): \textit{HoldsAt(k,t0+1)}, \textit{HoldsAt(g,t0+1)}
\item From $\Gamma_{t0}$ and \texttt{DEC5}: \textit{HoldsAt(f,t0+1)}
\end{enumerate}
Thus, $\Phi_{t0+1}(\Sigma, \Delta_{A_1(m_1)},\Psi_{1},\Gamma_{t0})$ = \{\textit{HoldsAt(k,t0+1)},  \textit{HoldsAt(g,t0+1)}, \textit{HoldsAt(f,t0+1)}\}.
\vspace{0.15cm}

\noindent Then, $\Phi_{t0+1}(\Sigma, \Delta_{A_2(m_2)},\Psi_{2},\Gamma_{t0})$ is calculated:

\begin{enumerate}\setcounter{enumi}{4}
\item From $\Sigma_{3}$,  $m_{2}\in A_{2}$, and due to the binding of the variables taking into account the current scenario: \textit{Initiates(send($A_2(m_2)$),k,t0)} 
\item From $\Delta_{A_2(m_2)}$, 5 and \texttt{DEC9}: \textit{HoldsAt(k,t0+1)}
\item From $\Gamma_{t0}$ and \texttt{DEC5}: \textit{HoldsAt(f,t0+1)}
\end{enumerate}
Thus, $\Phi_{t0+1}(\Sigma, \Delta_{A_2(m_2)},\Psi_{2},\Gamma_{t0})$=\{\textit{HoldsAt(k,t0+1)}, \textit{HoldsAt(f,t0+1)}\}. In this case, it can be concluded that $\Phi_{t0+1}(\Sigma, \Delta_{A_2(m_2)},\Psi_{2},\Gamma_{t0}) \not\models \Phi_{t0+1}(\Sigma, \Delta_{A_1(m_1)},\Psi_{1},$ $\Gamma_{t0})$.

However, if in the previous scenario the initial context $\Gamma_{t0}$ is modified to $\Gamma_{t0}=\emptyset$ then fluent $g$ is not initiated in step 2, and thus, it will not get to hold. So, in this case $\Phi_{t0+1}(\Sigma, \Delta_{A_1(m_1)},\Psi_{1},\Gamma_{t0})=\{\textit{HoldsAt(k,t0+1)}\}$, $\Phi_{t0+1}(\Sigma, \Delta_{A_2(m_2)},\Psi_{2},\Gamma_{t0})=\{\textit{HoldsAt(k,t0+1)}\}$ and as consequence:  $\Phi_{t0+1}(\Sigma,$ $ \Delta_{A_2(m_2)},\Psi_{2},\Gamma_{t0})\models \Phi_{t0+1}(\Sigma,$ $\Delta_{A_1(m_1)},\Psi_{1},\Gamma_{t0})$, which indicates that for the context $\Gamma_{t0}=\emptyset$, $m_2$ is a correct conversion for $m_1$.

\section{The proposed framework at work}
In this section one scenario that illustrates the different steps that are followed by the proposed framework to achieve semantic communication between software agents from two different information systems is presented. System \textsc{MedicalFIPAAgents} is composed of agents that use FIPA-ACL as agent communication language and FIPA-SL0 (\citep{FIPA-SL}) as content language at the time of composing messages. Moreover, system \textsc{Aingeru} is composed of agents that use specific communication acts in their messages (described in the applications layer of the \textsc{CommOnt} ontology). The format of the messages understood by the agents in \textsc{Aingeru} is the OWL format.

\begin{figure}
\begin{center}	
	\includegraphics[width=3.9in]{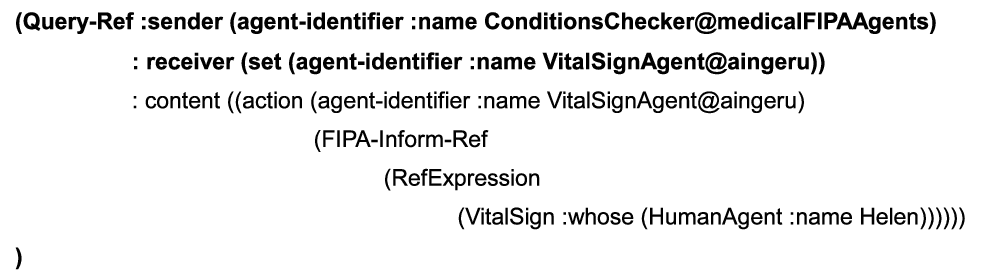}
	\caption{\label{scenarioMessage} A \textsc{MedicalFIPAAgents} message}
\end{center}
\end{figure}

In this example, the ConditionsChecker agent of system \textsc{MedicalFIPAAgents} wants to send the message in Fig.~\ref{scenarioMessage} to the VitalSignAgent in system \textsc{Aingeru} in order to make a request for the information concerning the vital signs of the patient Helen. 
%So, there will be two CommOntManagers (one per information system) that will create one CommOntManagerChild each, and each one with a different behaviour. 
%The CommOntManagerChild of the \textsc{MedicalFIPAAgents} system will follow \textit{Behaviour 1} described in section \ref{CMC}, while the CommOntManagerChild of the \textsc{Aingeru} system will follow \textit{Behaviour 2}.

When the CommOntManager of the system \textsc{MedicalFIPAAgents} receives the message from an agent of its own system, firstly, it transfers it to the TranslationMediator module, which decomposes it into two parts: the envelope and the content. This decomposition is performed taking into account the syntax of the message. In a FIPA-ACL message, the content is preceded by the tag ``:content", so it is easy to see that the content will be the text that goes from that tag to the following FIPA-ACL tag (or the end of the message). Fig.~\ref{scenarioMessage} shows both parts of the decomposed message (envelope in bold).

The following step consists on translating both submessages to the terms in the \textsc{CommOnt} ontology. In order to do so, a FIPA-ACL$\rightarrow$\textsc{CommOnt} translator will be used to translate the envelope of the message from FIPA-ACL to \textsc{CommOnt}, whereas a FIPA-SL0$\rightarrow$\textsc{CommOnt} translator will be used to translate the content of the message from FIPA-SL0 to \textsc{CommOnt}. Further details of these translators can be found in \citep{Berges11b}.

\begin{figure}
\begin{center}	
	\includegraphics[width=4.7in]{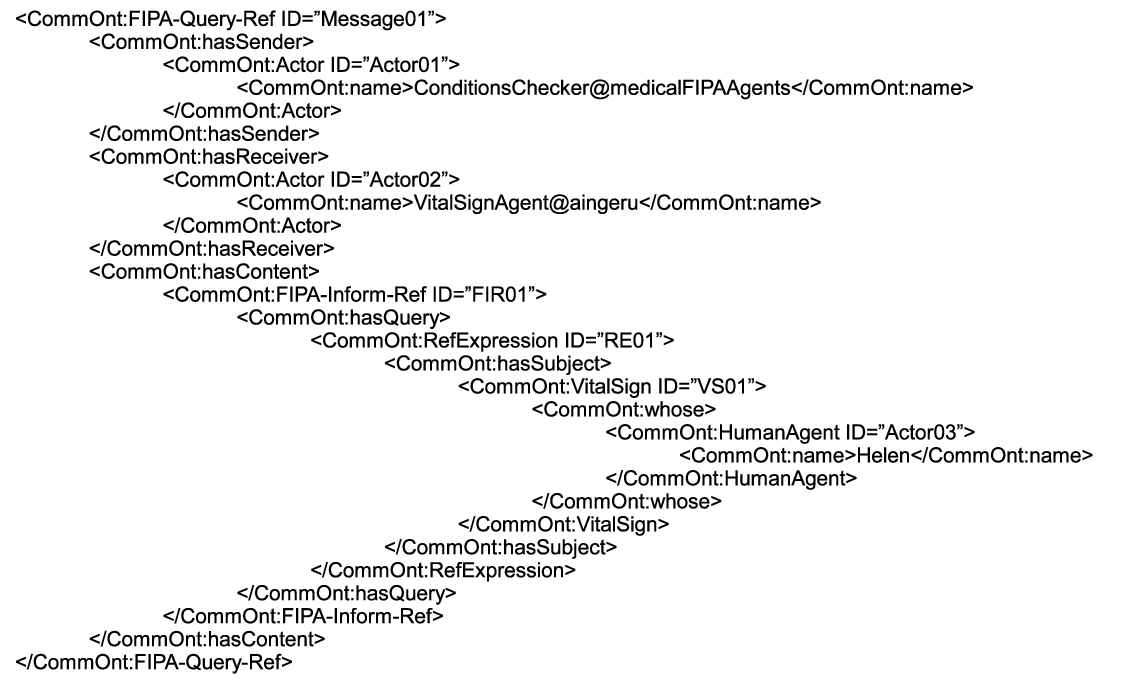}
	\caption{\label{scenarioCommOntFipa} Message in CommOnt}
\end{center}
\end{figure}

Then, both submessages are unified to create a complete message written in \textsc{CommOnt} (Fig.~\ref{scenarioCommOntFipa}) and given back to the CommOntManager. Next, the unified message is asserted\footnote{In the context of Description logics, asserted temporarily in the ABox for ABox realization.} into the \textsc{CommOnt} ontology and, with the help of a reasoner, entailments from the source message are collected (so that the information of the original message is made explicit). 
Some of the most relevant assertions, including entailed assertions, are shown next\footnote{For the sake of visual clarity, the namespace \texttt{CommOnt} is not included in the items that appear in the assertions.}:

\small
\begin{center}
\begin{tabular}{r @{  } c @{  } l|r @{  } c @{  } l}
\texttt{$\langle$Message01} & \texttt{rdf:type} & \texttt{FIPA-Query-Ref$\rangle$} & \texttt{$\langle$FIR01} & \texttt{rdf:type} & \texttt{ReportAct$\rangle$}\\
\texttt{$\langle$Message01} & \texttt{rdf:type} & \texttt{Inquiry$\rangle$} & \texttt{$\langle$FIR01} & \texttt{rdf:type} & \texttt{Content$\rangle$}\\
\texttt{$\langle$Message01} & \texttt{rdf:type} & \texttt{Directive$\rangle$} & \texttt{$\langle$FIR01} & \texttt{hasQuery} & \texttt{RE01$\rangle$}\\
\texttt{$\langle$Message01} & \texttt{hasModeOfAchiv} & \texttt{`polite'$\rangle$} & \texttt{$\langle$RE01} & \texttt{rdf:type} & \texttt{RefExpression$\rangle$}\\
\texttt{$\langle$Message01} & \texttt{hasContent} & \texttt{FIR01$\rangle$} & \texttt{$\langle$RE01} & \texttt{hasSubject} & \texttt{VS01$\rangle$}\\
\texttt{$\langle$FIR01} & \texttt{rdf:type} & \texttt{FIPA-Inform-Ref$\rangle$} & \texttt{$\langle$VS01} & \texttt{rdf:type} & \texttt{VitalSign$\rangle$}\\
\end{tabular}
\end{center}
\normalsize

Finally, the CommOntManager makes inquiries to the CommOntManagerAssistant\footnote{The CommOntManagerAssistant manages the information of existing CommOntManagers. When an information system incorporates its CommOntManager for the first time, this event must be shared with the CommOntManagers of the other systems, so that they are aware that it is possible to communicate with the new CommOntManager from that moment onward. } about the identity and location of the CommOntManager of the receiver's system (\textsc{Aingeru} system in this case) and sends the message to it. 

The set of OWL2 assertions sent by the CommOntManager of the system \textsc{MedicalFIPAAgents} is received by the CommOntManager of the \textsc{Aingeru} system. Then, the first step consists of asserting the set of axioms into the \textsc{CommOnt} ontology to infer new assertions that relate the terms of the original message with the ones understood by the agents of the \textsc{Aingeru} system. This will be possible thanks to the alignment axioms that have been defined in the \textsc{CommOnt} ontology between the terms in the applications layer and the terms in the common layer. 

For the sake of the example, let us imagine that the following terms are defined in the applications layer of the \textsc{CommOnt} ontology of the \textsc{Aingeru} system:

%\scriptsize
\begin{itemize}
\item\texttt{A-VitalSignQueryRef} $\equiv$ \texttt{Inquiry} $\sqcap$ $\exists$\texttt{hasContent.VitalSignInfGive} 
\item\texttt{VitalSignInfGive} $\equiv$ \texttt{ReportAct} $\sqcap$ $\exists$\texttt{hasQuery.VitalSignInfRef}
\item\texttt{VitalSignInfRef} $\equiv$ \texttt{RefExpression} $\sqcap\exists$\texttt{hasSubject.VitalSign}
\end{itemize}
\normalsize

Because of the assertions \texttt{$\langle$RE01} \texttt{rdf:type} \texttt{RefExpression$\rangle$}, \texttt{$\langle$RE01} \texttt{hasSubject} \texttt{VS01$\rangle$}, \texttt{$\langle$VS01} \texttt{rdf:type} \texttt{VitalSign$\rangle$} and the definition of the class \texttt{VitalSignInfRef}, the new assertion
\texttt{$\langle$RE01} \texttt{rdf:type}
 \texttt{VitalSignInfRef$\rangle$} is inferred.

Moreover, due to the assertions \texttt{$\langle$FIR01} \texttt{rdf:type} \texttt{ReportAct$\rangle$}, \texttt{$\langle$FIR01} \texttt{hasQuery} \texttt{RE01$\rangle$}, \texttt{$\langle$RE01} \texttt{rdf:type} \texttt{VitalSignInfRef$\rangle$}
and the definition of the class \texttt{VitalSignInfGive}, the new assertion \texttt{$\langle$FIR01} \texttt{rdf:type} \texttt{VitalSignInfGive$\rangle$} is inferred.

Finally, thanks to the assertions \texttt{$\langle$Message01} \texttt{rdf:type} \texttt{Inquiry$\rangle$}, \texttt{$\langle$Message01} \texttt{hasContent} \texttt{FIR01$\rangle$}, \texttt{$\langle$FIR01} \texttt{rdf:type} \texttt{VitalSignInfGive$\rangle$} and the definition of the class \texttt{A-VitalSignQueryRef}, the new assertion \texttt{$\langle$Message01} \texttt{rdf:type} \texttt{A-VitalSignQueryRef$\rangle$} is inferred. The same procedure is applied to the remaining assertions until all the new information is obtained.

The next step consists of writing the information represented by the new set of assertions into an OWL format message. 
\begin{figure}
\begin{center}	
	\includegraphics[width=4.8in]{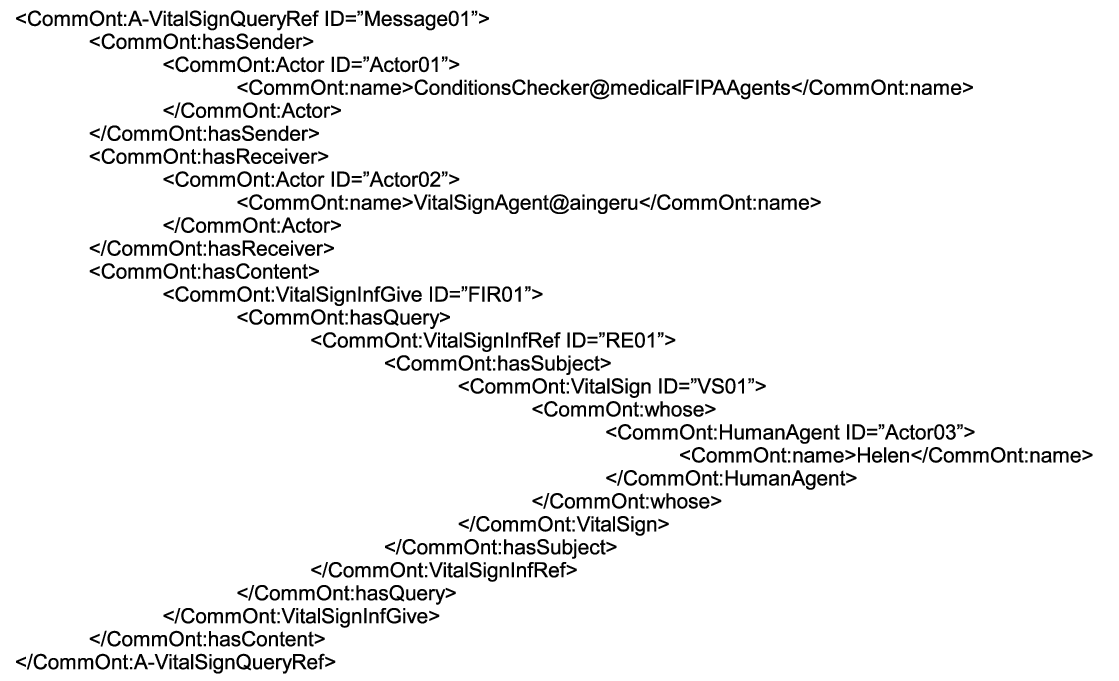}
	\caption{\label{scenarioFinalMessage} Message received by the recipient agent}
\end{center}
\end{figure}
In Fig.~\ref{scenarioFinalMessage} the generated message is shown. It can be observed that now the message contains individuals whose most direct classes (\texttt{A-VitalSignQueryRef}, \texttt{VitalSignInfGive}, \texttt{VitalSignInfRef}) are specific to the \textsc{Aingeru} system. In the last step, the message in the OWL format must be translated into a format understood by the agents of the recipient system. In order to do so, first the message must be decomposed into its two parts (the envelope and the content). In this case, this task is also an easy one because the content is located between the \textless \texttt{CommOnt:hasContent}\textgreater\ and \textless/\texttt{CommOnt:hasContent}\textgreater\ tags. Then again the use of some translators is required in order to perform the translation of both the envelope and the content (notice that in our specific example, these last two steps are avoidable, because the message format understood by the agents of the \textsc{Aingeru} system is the OWL format itself, so there is no need to use a translator). 
So, finally, the CommOntManager of the \textsc{Aingeru} system looks which specific agent of its system must be the receiver of the message, VitalSignAgent in this case, and it sends the message to it.
Next, a proof which shows that the conversion is satisfactory is presented.

\subsection*{\textbf{Proof: ``Conversion of Message01 is satisfactory"}}

For the sake of brevity, the names that appear in the example have been modified to the following:

\begin{center}
\begin{supertabular}{ll}
&\\
\textit{Fqr}=\textit{Fipa-Query-Ref} & \textit{Avsqr}=\textit{A-VitalSignQueryRef}\\
\textit{Fir}=\textit{Fipa-Inform-Ref} & \textit{Vsir}=\textit{VitalSignInfRef}\\
\textit{m}=\textit{Message01} & \textit{Vsig}=\textit{VitalSignInfGive}\\
\textit{a01}=\textit{Actor01} & \textit{a02}=\textit{Actor02}\\
\textit{f01}=\textit{FIR01}&\\
\end{supertabular}
\end{center}

\vspace{0.5cm}
\noindent We need to prove that $\Phi_{t0+1}$($\Sigma$,$\Delta_{Avsqr(m)}$,$\Psi_A$,$\Gamma_{t0}$)$\Rightarrow$
$\Phi_{t0+1}$($\Sigma$,$\Delta_{Fqr(m)}$,$\Psi_M$,$\Gamma_{t0}$) is valid, where $\Psi_A$ and $\Psi_M$ refer to the constraints of systems \textsc{Aingeru} and \textsc{MedicalFIPAAgents} respectively.

\newcommand{\indentitem}{\hspace{0.15in}}  

\noindent The following elements are common to both systems:

\begin{description}
\item[\indentitem(0)] $\Gamma_{t0}$=$\emptyset$
\item[\indentitem(1)] $\Sigma$=\{\textit{Initiates(send(Inquiry(?s,?r,?P))},\textit{CC(?r,?s,accept(?r,?s,?P),?P),t)}\}
\end{description}

\noindent For system \textsc{MedicalFIPAAgents} we know that
\begin{description}
\item[\indentitem(2)] $\Delta_{Fqr(m)}$=\{\textit{Happens(send(Fqr(m)),t0)}\}
\item[\indentitem(3)] $\Psi_{Fqr_1}, \Psi_{Fqr_2}\in\Psi_{M}$, where:
\item[\hspace{0.3in}(3.1)] $\Psi_{Fqr_1}$=\textit{Happens(send(Fqr(?m)),t)}$\rightarrow$ \textit{Happens(send(Inquiry(?m)),t)}
\item[\hspace{0.3in}(3.2)] $\Psi_{Fqr_2}$=\textit{HoldsAt(ReportAct(?r),t)} $\wedge$ \textit{HoldsAt(hasQuery(?r,?q),t)} $\wedge$ 
\item[\hspace{1.1in}]\textit{HoldsAt(RefExpression(?q),t)} $\rightarrow$ 
\textit{HoldsAt(Fir(?r),t)}
\end{description}
\noindent Now $\Phi_{t0+1}(\Sigma,\Delta_{Fqr(m)},\Psi_{M},\Gamma_{t0})$ is calculated:

\noindent From (2),(3.1) we get
\begin{description}
\item[\indentitem(4)] \textit{Happens(send(Inquiry(m)),t0)}
\end{description}
From binding variables in (1) with elements of \textit{m} we get 
\begin{description}
\item[\indentitem(5)] \textit{Initiates(send(Inquiry(a01,a02,f01)),CC1,t0)}, 
\item[\hspace{0.5in}]where \textit{CC1=CC(a02,a01,accept(a02,a01,f01),f01)}
\end{description}
From (4),(5) and \texttt{DEC9} we get
\begin{description}
\item[\indentitem(6)]  \textit{HoldsAt(CC1,t0+1)}
\end{description}
From the assertions in the source system (see Fig.~\ref{scenarioCommOntFipa}) about the content of \textit{m} and the law of inertia we get
\begin{description}
\item[\indentitem(7)] \textit{HoldsAt(Fir(f01),t0+1)}
\item[\indentitem(8)] \textit{HoldsAt(hasQuery(f01,RE01),t0+1)}
\item[\indentitem(9)] \textit{HoldsAt(RefExpression(RE01),t0+1)}
\end{description}

\noindent Then, $\Phi_{t0+1}(\Sigma,\Delta_{Fqr(m)},\Psi_{M},\Gamma_{t0})$
=\{(6),(7),(8),(9)\}

\vspace{0.5cm}
\noindent From system \textsc{Aingeru} we know that

\begin{description}
\item[\indentitem(10)] $\Delta_{Avsqr(m)}$=\textit{Happens(send(Avsqr(m)),t0)}
\item[\indentitem(11)] $\Psi_{Avsqr_1}, \Psi_{Avsqr_2}, \Psi_{Avsqr_3}\in\Psi_{A}$, where:
\item[\hspace{0.3in}(11.1)] $\Psi_{Avsqr_1}$=\textit{Happens(send(Avsqr(?m)),t)}$\rightarrow$
\item[\hspace{1.4in}]\textit{Happens(send(Inquiry(?m)),t)}
\item[\hspace{0.3in}(11.2)] $\Psi_{Avsqr_2}$=\textit{HoldsAt(Vsig(?v),t)}$\rightarrow$\textit{HoldsAt(ReportAct(?v),t)}
\item[\hspace{0.3in}(11.3)] $\Psi_{Avsqr_3}=$\textit{HoldsAt(Vsir(?v),t)}$\rightarrow$\textit{HoldsAt(RefExpression(?v),t)}
\end{description}

\noindent Now $\Phi_{t0+1}(\Sigma,\Delta_{Avsqr(m)},\Psi_{A},\Gamma_{t0})$ is calculated:

\noindent From (10),(11.1) we get 
\begin{description}
\item[\indentitem(12)] \textit{Happens(send(Inquiry(m),t0))}
\end{description}
From binding variables in (1) with elements of \textit{m} we get
\begin{description}
\item[\indentitem(13)] \textit{Initiates(send(Inquiry(a01,a02,f01)),CC1,t0)},
\item[\hspace{0.5in}]where \textit{CC1=CC(a02,a01,accept(a02,a01,f01),f01)}
\end{description}
From (12),(13) and \texttt{DEC9} we get 
\begin{description}
\item[\indentitem(14)] \textit{HoldsAt(CC1,t0+1)}
\end{description}
From the assertions in the \textsc{Aingeru} system (see Fig.~\ref{scenarioFinalMessage}) about the components of \textit{m} and the law of inertia we get
\begin{description}
\item[\indentitem(15)] \textit{HoldsAt(Vsig(f01),t0+1)}
\item[\indentitem(16)] \textit{HoldsAt(hasQuery(f01,RE01),t0+1)}
\item[\indentitem(17)] \textit{HoldsAt(Vsir(RE01),t0+1)}
\end{description}

\noindent Then, $\Phi_{t0+1}(\Sigma,\Delta_{Avsqr(m)},\Psi_{A},\Gamma_{t0})$
=\{(14),(15),(16),(17)\}
\begin{description}
\item[ ]
\end{description}
\noindent Once the set of observations at timepoint \textit{t0+1} has been calculated for each system, and considering that \{(14),(15),(16),(17)\} is glaringly consistent, we have to prove that \{(14),(15),(16),(17)\}$\Rightarrow$\{(6),(7),(8),(9)\}.

\noindent Notice that (14)=(6) and (16)=(8). From (15),(11.2) we get
\begin{description}
\item[\indentitem(18)]  \textit{HoldsAt(ReportAct(f01),t0+1)}
\end{description}
From (17),(11.3) we get
\begin{description}
\item[\indentitem(19)]  \textit{HoldsAt(RefExpression(RE01),t0+1)}, and (19)=(9)
\end{description}
From (18),(16),(19),(3.2) we get
\begin{description}
\item[\indentitem(20)]  \textit{HoldsAt(Fir(f01),t0+1)}, and (20)=(7)
\end{description}

\noindent and the proof is completed.
%\noindent So, in summary we have that
%\begin{description}
%\item[](14)$\rightarrow$(7)
%\item[](16)$\rightarrow$(9)
%\item[](19)$\rightarrow$(10)
%\item[](20)$\rightarrow$(8)
%\end{description}

%which justifies that $\Phi_{t1}(\Sigma, \Delta_{Avsqr(m)}, \Psi_{Avsqr(m)}, \Gamma_{t0})\models\Phi_{t1}(\Sigma, \Delta_{Fqr(m)}, \Psi_{Fqr(m)}, \Gamma_{t0})$

It can be concluded that, thanks to the proposed framework, two agents from different systems, which previously could not understand each other, have been able to communicate with one another without establishing a priori agreement on the format and vocabulary used in the interchanged messages.

\section{Related works}\label{Related}
Several works that treat the topic of interoperability among agents from heterogeneous systems can be found in the specialized literature.
The work of \citep{Lopes05} suggests the idea of translating the content of every message to an abstract logic language defined by the authors (and vice versa). This work is similar to ours because it makes use of an intermediate language. Nevertheless, it only takes into account the syntax of the messages, so it allows interoperability between messages with different formatting but does not address the problem from a semantic point of view. In our case, however, semantic interoperability is considered.

A different approach for interoperability, based on the inclusion of preformatted message templates within the advertised capability description of
agents is presented in \citep{Payne02}. The shallow-parsing template approach presented in that paper relaxes
the constraint that states that agents need to share a common language
for describing the content and format of messages.
However, it reduces the problem to sharing common
templates and as a side effect imposes certain
restrictions on the type of messages that can be
exchanged. From our point of view, a main
disadvantage is its heavy emphasis on syntactic
aspects.

Mechanisms that translate communications from
one multiagent system to another have been developed.
For instance, \citep{Giampapa00} describes an InterOperator
that implements a connection between the RETSINA
system (a KQML-based system) and the OAA system.
The Open Agent Architecture (OAA) \citep{Martin99} is a
framework for constructing multiagent systems
and their designers intended to minimize the effort
involved in wrapping legacy applications. Agent
communications are represented as events. Each
event has a type, a set of parameters, and a content.
The allowable content and parameters vary according
to the type of event. In our opinion the great
effort needed to implement this ad hoc interoperation
undermines the scalability of the approach.

The role of the \textsc{CommOnt} ontology explained in this paper is similar to 
the role of the PSL (Process Specification Language) ontology 
in~\citep{Gruninger05}, but we complement 
the system with an Event Calculus axiomatization of the framework in order to 
be able to sanction the satisfaction on the conversion of messages. Another work 
that takes into account the conversion problem is~\citep{Dou03}, where the 
first-order logic language Web-PDDL is used as a common representation language and 
a dedicated reasoner OntoEngine is used as inference mechanism. However, the fact that OntoEngine is not a complete reasoner and 
that Web-PDDL is a proprietary language, make a difference with our work that use  existing complete reasoners for the 
standard language OWL2-DL. Moreover, authors in \citep{Pasha06} propose the use of OWL ontologies to allow an effective and bidirectional communication between FIPA-compliant software agents and OWL-based Web Services. More precisely, they have defined a series of alignment axioms between the standard content language FIPA-SL and its representation in OWL, so that the latter can be used by Web Services. In \citep{Pasha10} their work is extended in order to allow negociation between agents and web services. However, this solution is centered only on one content language, so it needs to be developed more widely to achieve any real interoperability.

Another set of works deals with the semantics of agent communication languages. \citep{Boella09} introduces a semantics based on social networks. It proposes a semantics of speech acts as `relationship building', where the shape of those relationships are those usually found in social networks (e.g.\ dependency, authority, commitment). In other words, in this approach the pre- and post-conditions of speech acts are expressed as constraints on social networks. Authors in \citep{Boella06} present a role-based semantics for ACLs, understanding the notion of role as the description of an expected behaviour. The proposed approach embeds both the mental and social semantics approaches. On the one hand, the mentalistic approach is embedded by attributing mental attitudes to the public roles agents play instead of agents themselves. That is, it moves from the agents' beliefs and goals to the roles' beliefs and goals. On the other hand, the social approach is embedded by showing how roles maintain the normative character of social semantics. Finally, the work in \citep{Gaudou06} presents a semantics for FIPA-ACL, based on social attitudes. The private attitudes of mentalistic approaches are replaced by attitudes that are made public through communication. It also differs from the social commitments approach, since the state of the commitment of the participating agents is obtained directly from the logical post-conditions of the communication acts (i.e.\ publicly expressed intentions) instead of from what has been communicated and the commitments the agents have made by doing that. 

Recent research~\citep{Chopra12} has pointed out the 
consensus on using social abstractions for the specification of semantics for communication acts, instead of mentalist abstractions. 
Works such as~\citep{Fornara07, Singh00}
use social commitments to express that kind of semantics. Moreover, the Event 
Calculus~\citep{Mueller06} has been used to represent and to reason about such 
commitments~\citep{Yolum04, 
Fornara09}. In our work we also deal with social commitments and the Event Calculus.
However, the interesting difference 
between those works and our is that we focus on a different problem. Those works are concerned with the clean specification of semantics 
for communication acts that are assumed to be shared by the agents in the 
conversation; however, we are concerned with the problem of facilitating the 
communication to agents that eventually use different communication languages 
that must be somehow aligned and in  evaluating the comprehension 
of the exchanged messages.

To the best of our knowledge, there are few works considering the Semantic Web technology for the representation of agent communication. A remarkable exception 
is~\citep{Fornara10} 
where an OWL2 ontology is developed for representing commitments and their related 
items. The proposed ontology is designed with the goal of exploiting OWL reasoners 
in the run-time monitoring of the communication exchanged in the context of artificial 
institutions. Such reasoning can be considered a technical alternative to the 
reasoning procedures offered by the Event Calculus reasoners. Nevertheless, the 
goal of that work is to evaluate the compliance of the communicating agents to the 
norms and obligations of an artificial institution; and therefore is complementary 
to our work in this paper that is focussed on evaluating the comprehension 
of the exchanged messages. 

Finally, the closest work to ours is that of \citep{Chopra08}, which also proposes commitment-based interoperability. In particular, they consider constitutive interoperability, which takes into account only the meaning of messages, opposed to regulative interoperability, which takes into consideration message order, occurrence and data flow. Our work considers also this high-level definition of interoperability that takes into account the business meaning of communication, but it differs from theirs in the decision procedure for determining interoperability of pairs of agents.

\section{Conclusions}
We have introduced a proposal, based on Semantic Web technology, to tackle the problem of semantic communication among heterogeneous and distributed information systems represented by software agents. A novel feature of that proposal is that it favours a flexible communication process among those systems avoiding \textit{a priori} agreements about interchanged messages. Such flexibility is obtained by considering the semantics associated to the communication process. The proposed framework includes as central element an ontology whose main category represents communication acts according to the well-recognized speech acts theory. This ontology integrates, using alignment axioms, descriptions of classes of messages from different ACL standards as well as descriptions of non standard classes of messages from particular information systems. Alignment axioms in that ontology provide the setting for the conversion of a message from one system into a message for another system.
% Moreover, ontology reasoning based on message class descriptions enhances the processing performed by the translators. 
%On the other hand, a translation mediator that can incorporate reasoning in the translation process, which gives the chance to take advantage of the semantic information.
We have explained the necessary stages to be accomplished in a communication process in order to present the whole task to be solved. Finally, we have developed a first-order logic interpretation of the scenario and we have summarized the conversion context domain by presenting a guide for a proper axiomatization, which has been used to formalize the notion of satisfactory conversion of messages. The evaluation of whether a conversion has been satisfactory opens the doors to solving a bigger problem to be addressed in the future: assessing the quality of a conversion. Under this title we refer
to the idea of informing the participants about the degree of understanding
that has been achieved in a communication, that is to say, if the understanding
has been total or partial, and in the latter case, suggesting some
possible changes to facilitate understanding. Previous works of members of
our research group in the area of estimation of information loss for multiontology
based query processing \citep{Mena00} will be a good start point to tackle
the considered problem.



%% References


%\bibliographystyle{elsarticle-num}






% \end{thebibliography}
\section*{Acknowledgements}
\noindent{T}he work of {I}doia {B}erges was supported by a grant of the {B}asque {G}overnment ({P}rograma de {F}ormaci\'on
de {I}nvestigadores del {D}epartamento de {E}ducaci\'on, {U}niversidades
e {I}nvestigaci\'on). {T}his work is also supported the Spanish Ministry of Education and Science {TIN}2010-21387-{C}02-01.

\bibliographystyle{elsarticle-harv}

%\bibliography{BDI}

\newpage
\appendix

\section{}\label{appendix:stateMachines}

\begin{figure}[h]
\begin{center}	
	\includegraphics[width=4.9in]{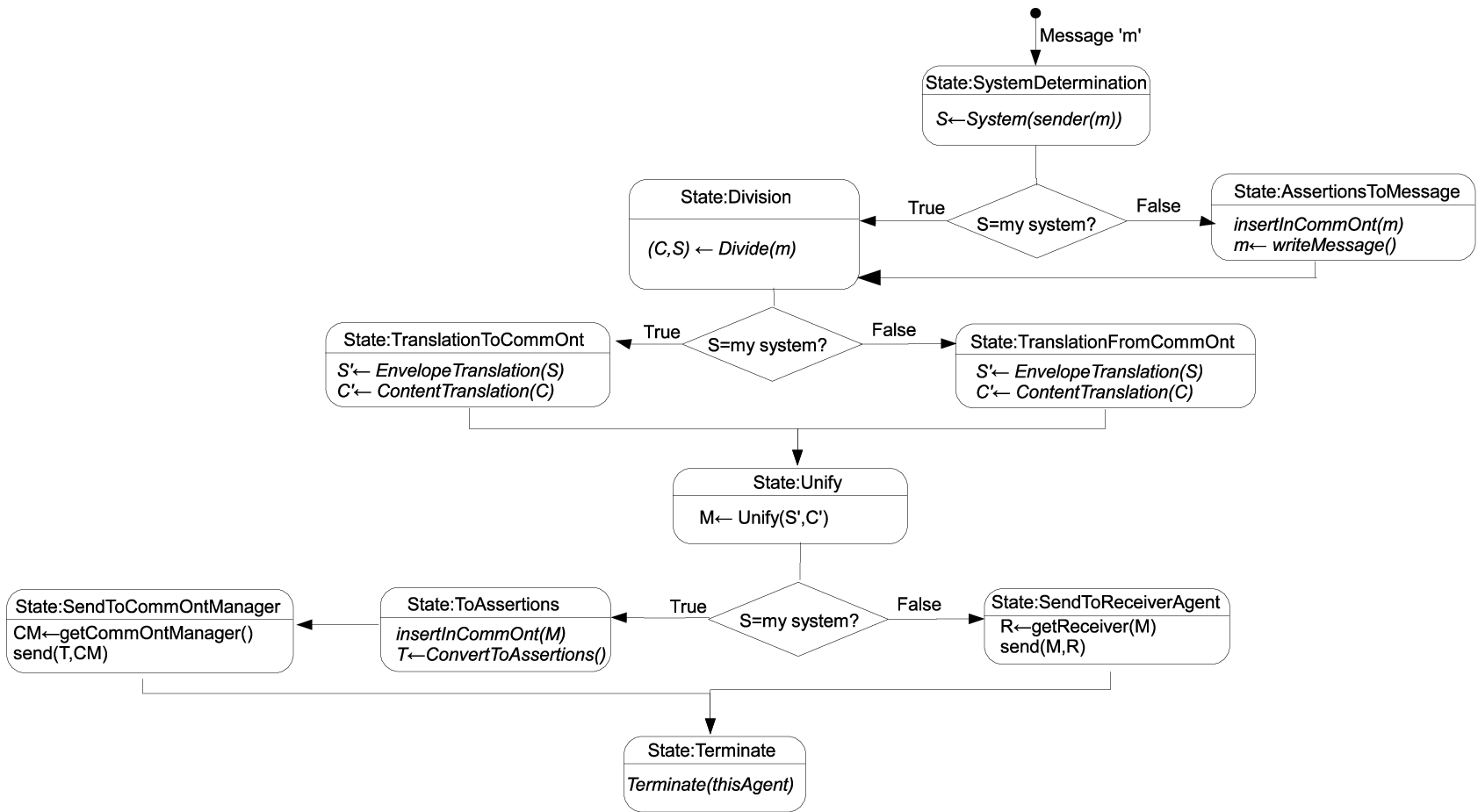}
	\caption{\label{CMstateMachine} State machine of a CommOntManager agent}
\end{center}
\end{figure}

\end{document}